\documentclass[runningheads]{llncs}

\usepackage[T1]{fontenc}
\usepackage{graphicx}
\usepackage{booktabs}
\usepackage{array}
\usepackage{multirow}
\usepackage{amsmath}
\usepackage[hypertexnames=false]{hyperref}
\usepackage{color}

\usepackage{url}
\usepackage{float}
\usepackage{paralist}
\usepackage{tikz}
\usetikzlibrary{positioning,arrows.meta,calc,fit,backgrounds}

\usepackage[normalem]{ulem}
\usepackage[colorinlistoftodos]{todonotes}
\usepackage[colorinlistoftodos]{todonotes}
\usepackage{xcolor}
\usepackage[draft,noindex]{fixme}
\fxsetup{theme=color,mode=multiuser}
\FXRegisterAuthor{ah}{aah}{{\color{magenta}Anett Hoppe}}
\FXRegisterAuthor{rs}{ars}{{\color{blue}Ratan Sebastian}}
\FXRegisterAuthor{ee}{aee}{Elias Entrup}

\begin{document}

\title{Using Hierarchical Controlled Vocabularies to Understand CLIP Retrieval Failures in Historical Photo Collections}

\titlerunning{Using Hierarchical Controlled Vocabularies to Understand CLIP Failures}

\author{Ratan Sebastian\inst{1,2} \and
Anett Hoppe\inst{3,4,1} \and
Christoph Rippe\inst{5} \and
Ralph Ewerth\inst{3,4,1,2}}
\authorrunning{Sebastian et al.}
\institute{
TIB -- Leibniz Information Centre for Science and Technology, Hannover, Germany \and
L3S Research Center, Leibniz University Hannover, Germany \and
Marburg University, Germany \and
Hessian Center for Artificial Intelligence (hessian.AI), Marburg/Darmstadt, Germany \and
Goethe University Frankfurt, University Library Frankfurt, Germany}

\maketitle

\begin{abstract}
GLAM institutions (Galleries, Libraries, Archives, and Museums) organise image access using controlled vocabularies such as the Art and Architecture Thesaurus (AAT).
For content-based image retrieval in these settings, vision-language models like CLIP are increasingly used, but their performance varies.
This variation is known to relate to measures like concept abstraction and concept frequency.
However, no prior work explains this variation in terms of the structural properties of vocabularies like the AAT that GLAM professionals already use.
The AAT groups concepts into broad facets (\emph{Objects}, \emph{Activities}, \emph{Agents}, etc.) and arranges terms hierarchically within them.
In this paper, we ask whether two structural properties (root facet type and hierarchy depth) explain where CLIP retrieval succeeds and fails, and where fine-tuning helps.
Across three historical photographic collections annotated with AAT terms, we examine visual coherence (whether a term's photographs cluster in CLIP's embedding space), text-image alignment (whether its label is near that cluster), and standard retrieval measures, which conflate the two.
We find that visual coherence and text-image alignment are nearly uncorrelated across terms and jointly separate distinct failure modes.
Terms whose photographs cluster tightly but whose label is distant from the cluster retrieve poorly in every collection, in two of three collections even worse than terms that fail on both metrics.
We also show that while retrieval metrics do not correlate significantly with either structural property, root facet type does significantly separate categories with varying visual coherence.
Finally, we find that fine-tuning improves retrieval overall, but its gains favour shallower terms in the hierarchy, where text-image alignment improves most, beyond what concept frequency explains.
Together, these results make CLIP's failure more interpretable through the structure of a domain-specific vocabulary like the AAT.

\keywords{CLIP \and image retrieval \and visual coherence \and Art and Architecture
Thesaurus \and historical photographic collections \and domain adaptation}
\end{abstract}

\section{Introduction}
\label{sec:intro}

GLAM institutions are deploying vision-language models like CLIP (Contrastive Language-Image Pre-training) for content-based image retrieval in photographic collections~\cite{smits2023}.
Unlike general web image retrieval, archival retrieval is organised around controlled vocabularies~\cite{late2024}.
GLAM professionals assign controlled vocabulary terms to describe the content of photographs, and users rely on these terms as their primary search entry points~\cite{late2024}.
CLIP enables zero-shot retrieval (querying without collection-specific training data), which is especially attractive in GLAM institutions where metadata is often incomplete, inconsistent, or missing entirely~\cite{smits2023,aske2023}.
But the domain shift between CLIP's training domain which is mostly general web images and the historical archives domain leads to issues like the model not understanding domain specific terms or visual concepts~\cite{aissi2025}.

Previous work has shown that CLIP performance varies by concept abstraction level~\cite{rodriguez2024,verma2023}. 
Conceptual queries such as \textit{family vacations} or \textit{working class life} retrieve far fewer relevant results than descriptive ones such as \textit{people chopping vegetables}~\cite{rodriguez2024}.
It also varies by concept frequency~\cite{parashar2024,alijani2026}. 
Low-frequency terms such as \textit{battle of the bulge}, \textit{war bonds}, and \textit{sailboats} are retrieved poorly by zero-shot CLIP as opposed to high frequency ones such as \textit{aircraft}, \textit{medical service} and \textit{airstrike} in historical photographic collections~\cite{alijani2026}.
These accounts explain failure in terms of WordNet position~\cite{rodriguez2024} and pretraining corpus frequency~\cite{alijani2026}.
But no work so far has captured its variation in terms of a controlled vocabulary such as the Art and Architecture Thesaurus (AAT)~\cite{aat2024} that GLAM professionals already use to organise image access.
Further, previous work has characterized CLIP failure only in terms of retrieval metrics such as Recall@K and NDCG@K (Normalised Discounted Cumulative Gain), which confound whether photographs in a category cluster visually and whether the category label is recognised by CLIP's text encoder.

In this paper, we address these two gaps.
We use the AAT, which organises concepts into broad facets and arranges terms hierarchically within them, to correlate CLIP behaviour (measured by visual coherence, text-image alignment and retrieval performance) with two structural properties (\textbf{root facet type} and \textbf{hierarchy depth}) across three historical photographic archives.
Following Leemann et al.~\cite{leemann2022}, we measure intra-category \textbf{visual coherence} ($d_{\text{mean}}$): the mean pairwise cosine similarity of CLIP image embeddings within a vocabulary category.
A high value means photographs cluster tightly in CLIP's visual space.
Separately, we measure \textbf{text-image alignment}: the cosine similarity between the CLIP text embedding of the category label and the centroid of its image embeddings.
A high value means the label lands near the image cluster.
Their joint distribution identifies three distinct sources of CLIP retrieval failure, detailed in Section~\ref{sec:failure}: photographs that are visually incoherent, labels that land far from a coherent cluster, and labels that land near a cluster too diffuse for precise retrieval.
The distinction matters because it lets us ask which of these sources of difficulty fine-tuning strategies actually repair.
We find, for instance, that fine-tuning influences mainly text-image alignment and not visual coherence.
\textbf{Root facet types} are the AAT's~\cite{aat2024} major concept classes which are ordered from abstract to concrete: \emph{Associated Concepts} (abstract ideas), \emph{Styles and Periods} (art historical styles and periods), \emph{Agents} (people and organizations), \emph{Activities} (actions and processes), \emph{Materials} (physical substances), and \emph{Objects} (tangible, human-made things).
\textbf{Hierarchy depth} measures how many levels below the root facet a term sits: shallower terms name broad categories, deeper terms name specific concepts that are a kind of their broader context.
These two properties correspond to the concreteness and specificity dimensions along which CLIP performance is known to vary, as discussed earlier in this section.
To understand the stability of our results across photographic collections we apply our analysis to three of them from different contexts: colonial-era German photos~\cite{vabiko}), early 20th-c.\ Catalan press photos~\cite{sagarra2024}, and FAIR Photos~\cite{vanwissen2024}  which contain postwar Dutch press photos.

Concretely, our research questions are:
\begin{description}
\item[RQ1] Do AAT \textbf{root facet type} and \textbf{hierarchy depth} correlate with \textbf{visual coherence}, \textbf{text-image alignment}, and standard retrieval metrics (Recall@$K$ and NDCG@$K$)?
\item[RQ2] Do AAT \textbf{root facet type} and \textbf{hierarchy depth} correlate with fine-tuning improvement?
\end{description}

The following are some key findings from our study.
The correlation between \textbf{visual coherence} and \textbf{text-image alignment} is close to zero, both across all collections and within each collection, meaning that these two factors that lead to retrieval failures are independent.
For example, in the \emph{Materials} facet photographs cluster tightly, yet labels are the worst-aligned.
This independence property helps us decompose retrieval failure and fine-tuning successes to explain which of these two dimensions is most affected.
We can see, for example, that in the zero-shot setting the structural properties of AAT terms are more correlated with \textbf{visual coherence} than overall retrieval performance.
For instance, deeper AAT terms form tighter image clusters than shallow, broad ones, a relationship that holds up across collections, yet this same ranking by \textbf{hierarchy depth} shows essentially no relationship to NDCG@10. 
Interestingly, we see that terms whose photographs cluster tightly but whose label is far from that cluster retrieve worse than terms that are both diffuse and poorly aligned.
Also, as previously stated, we can see that fine-tuning mainly improves text-image alignment, not visual coherence.

\section{Related Work}
\label{sec:relwork}

We review CLIP-based retrieval in cultural heritage settings, image indexing research that grounds the AAT's structural properties as predictors of retrieval behaviour, and accounts of CLIP's known limits, which our work extends from linguistic and data properties to vocabulary structure.

\subsection{CLIP in GLAM}

Multimodal retrieval is changing how digital humanities research engages with photographic collections, enabling exploration and analysis at scales previously impractical~\cite{smits2023}.
CLIP retrieves via text queries without collection-specific training, letting GLAM institutions with incomplete, inconsistent, or absent metadata make collections discoverable without manual cataloguing~\cite{aske2023}.
Deploying generic CLIP in archival settings still has challenges.
The visual domain shift between web-trained features and historical photographic material causes issues: halftone printing, physical degradation, and microfilm intermediaries produce visual characteristics generic CLIP has not encountered, leading to poor retrieval~\cite{aissi2025}.
Specialised domain vocabulary also creates text-image alignment failures~\cite{yuan2025}, and fine-grained categories require representations generic CLIP does not provide~\cite{balauca2024}.

Fine-tuning on domain-specific data addresses these problems: adapting CLIP to historical archives with long-tailed distributions improves retrieval, with progressive unfreezing raising Recall@1 from 15\% to 68\%~\cite{alijani2026}.
It remains unclear, however, whether these gains are uniform across low-frequency categories or concentrated in types meaningful to collection users, the gap this work addresses.

\subsection{Controlled vocabularies for images}

Image indexing in library and information science (LIS) organises image content along two dimensions: ontological type (what kind of thing is depicted) and abstraction level (how far a description is from a concrete visual referent).
This understanding derives from three influential frameworks: Jörgensen's Pyramid of Visual Descriptors~\cite{jorgensen2001}, which takes ontological category as its primary axis; the Visual Information Vocabulary~\cite{jorgensen2005}, which formalises this into facets such as Natural Objects, Living Beings, and Abstract Concepts; and Westman~\cite{westman2009}, who approaches the same space from the user side, categorizing information needs as generic, specific, or abstract, and further into objects and activities.
\emph{Objects} have stable visual forms while \emph{Activities} and abstract concepts do not.

The AAT~\cite{aat2024} is an authority file maintained by the Getty Vocabulary Program, used at cataloguing time across GLAM institutions to assign controlled terms to collection objects, including photographs, promoting consistency and cross-collection retrieval.
Its structure maps onto both dimensions identified above: its seven root facets (introduced in Section~\ref{sec:intro}) encode ontological type, ordered from abstract to concrete, and within each facet the hierarchy encodes genus/species relationships, so depth measures conceptual specificity.

\subsection{Limits of CLIP}

CLIP~\cite{radford2021} learns joint image-text representations from web image-caption pairs via a contrastive objective, retrieving via cosine similarity between a text query and image embeddings.
Previous work shows a fivefold drop in Recall@1 from caption-style to conceptual queries~\cite{rodriguez2024}, with WordNet hyponymy depth as the primary predictor: concepts closer to the WordNet root are harder to retrieve.
Concept frequency also affects retrieval performance: concepts underrepresented in web training data are retrieved less accurately, regardless of abstraction level~\cite{parashar2024,alijani2026}.
Term visualness (whether a word evokes a mental image) predicts retrieval quality~\cite{verma2023}, and category visual coherence (the degree to which a category's photographs cluster in CLIP's embedding space) correlates with human judgements of concept coherence~\cite{leemann2022}.

Across these accounts, CLIP failure is explained by linguistic properties (WordNet depth, visualness) or data properties (concept frequency), never by the structural properties of a vocabulary GLAM institutions already use, such as the AAT.
We do so here: if CLIP's success and failure correlate with the AAT's structural divisions, GLAM professionals gain a way to anticipate retrieval problems using a framework they already understand.

\section{Datasets}

We study three historical photographic collections, each annotated with AAT terms and spanning different geographic regions, periods, and subject matter. All three are black-and-white photographs but differ in era, origin, and digitisation practice, so findings that hold across all three are unlikely to be artefacts of any one collection. Each is described below and summarised in Table~\ref{tab:datasets}.

\subsection{University Library Frankfurt Collection from Colonial Contexts}
\label{sec:GermanColonial}

This dataset draws from the Image Archive from Colonial Contexts, University Library Frankfurt~\cite{vabiko} (henceforth abbreviated \emph{GermanColonial}).
The collection contains over~50{,}000 photographs of the German Colonial Society and subsequent colonial organisations.
We created AAT links by mapping German content keywords from the original metadata to AAT term IDs.
The metadata contains 8{,}124 unique keywords across 62{,}519 images.
Matching proceeds in two stages:
\begin{inparaenum}[(1)]
\item exact case-insensitive matching against the AAT (193 keywords matched, covering 26.4\% of images);
\item spaCy German lemmatization (\texttt{de\_core\_news\_lg}), which maps inflected forms to their canonical base form and allows matching across morphological variants.
\end{inparaenum}
Together the two stages yield 1{,}429 matched keywords (17.6\% of unique keywords), covering 35{,}159 images (56.2\% of the archive). The remaining 43.8\% have no AAT match and are excluded.
A manual check of a random sample of 50 matched keywords identified only 3 errors: one instance of metadata noise (keywords such as \textit{s}) and two instances of ambiguous matches (e.g.\ \textit{Schiff} matching both \textit{ship} and \textit{nave} when only one sense applies).
While we know that the mapping procedure is not perfect, the low error rate is sufficient to test correlations across hundreds of keywords.
The final linked datasets is dominated by \emph{Objects} which comprise 81\% of terms (Table~\ref{tab:facet_dist}).

\subsection{Photographs from Catalan Photographer Josep Maria Sagarra}

The Sagarra dataset~\cite{sagarra2024} comprises 8{,}536 historical photographs taken by Josep Maria Sagarra, one of the most prominent Catalan press photographers of the first half of the twentieth century, held by the Ajuntament de Girona and retrieved via the Europeana API(henceforth abbreviated \emph{Sagarra}).
AAT identifiers are embedded directly in the Europeana metadata as Getty vocabulary URIs, requiring no mapping step and yielding near-complete coverage (99.7\%).
The 111 analysed terms span a more balanced facet distribution than GermanColonial, with a higher proportion of \emph{Agents} (16\%) and \emph{Activities} (11\%).

\subsection{FAIRPhotos}

FAIR Photos~\cite{vanwissen2024} is a cross-institutional collection of 18{,}143 historical photographs aggregated from multiple European cultural heritage repositories under open licenses.
As with Sagarra, AAT identifiers originate from Getty vocabulary URIs embedded in the source metadata.
FAIR Photos has the most diverse facet distribution of the three archives with \emph{Activities} accounting for 40\% of terms which is comparable to \emph{Objects} (38\%) making it a useful contrast to GermanColonial's object-dominated vocabulary.

\begin{table}[t]
\caption{Collection overview and AAT root facet distribution. AAT terms are those with $\geq 20$ images in that collection (Section~\ref{sec:method}). ``---'' $\Rightarrow$ facet absent.}
\label{tab:datasets}\label{tab:facet_dist}
\centering\small
\setlength{\tabcolsep}{5pt}
\begin{tabular}{lrrr}
\toprule
 & GermanColonial & Sagarra & FAIR Photos \\
\midrule
Photographs     & 34{,}882 &  6{,}274 & 17{,}605 \\
AAT terms       & 378 & 111 & 85 \\
Depth (median)  & 7 & 7 & 6 \\
Depth range     & 3--13 & 3--13 & 1--11 \\
\midrule
\multicolumn{4}{c}{\textit{Facet distribution (\% of terms)}} \\[2pt]
\midrule
\emph{Objects} & 81\% & 70\% & 38\% \\
\emph{Agents} & 8\% & 16\% & 11\% \\
\emph{Activities} & 2\% & 11\% & 40\% \\
\emph{Assoc.\ Concepts} & 3\% & 3\% & 9\% \\
\emph{Materials} & 3\% & --- & 2\% \\
\emph{Styles \& Periods} & 2\% & --- & --- \\
\emph{Physical Attributes} & 0\% & --- & --- \\
\bottomrule
\end{tabular}
\end{table}

\section{Design of CLIP Analysis based on Vocabulary Structure}
\label{sec:method}

\begin{figure}[tbp]
\centering
\resizebox{0.7\textwidth}{!}{%
\begin{tikzpicture}[
  font=\small,
  box/.style          = {draw, rectangle, align=center, minimum width=36mm,
                          minimum height=20mm, inner sep=3pt, line width=0.5pt},
  modelbox/.style     = {draw, rectangle, align=center, minimum width=36mm,
                          minimum height=12mm, inner sep=2pt, line width=0.5pt},
  metricsbox/.style   = {draw, rectangle, align=center, minimum width=32mm,
                          minimum height=38mm, inner sep=4pt, text width=36mm,
                          line width=0.5pt},
  statsbox/.style     = {draw, rectangle, align=left,   minimum width=46mm,
                          minimum height=38mm, inner sep=4pt, text width=46mm,
                          line width=0.5pt},
  arr/.style  = {-{Latex[length=1.6mm, width=1.4mm]}, line width=0.5pt},
  edge/.style = {line width=0.5pt},
  groupbox/.style = {draw=gray!55, dashed, line width=0.4pt,
                      rounded corners=2pt, inner sep=7pt},
  grouplbl/.style = {font=\scriptsize\scshape, color=gray!70,
                      anchor=south west, inner sep=1pt},
]


\node[box] (collections) at (0, 1.2) {%
  \textbf{Collections}\\[1pt]
  GermanColonial\\ Sagarra\\ FAIR Photos
};

\node[box] (aat) at (0, -1.2) {%
  \textbf{AAT properties}\\[1pt]
  Root facet\\ Depth
};

\node[modelbox] (lora) at (5.6,  1.5) {%
  \textbf{LoRA}\\
  {\scriptsize fine-tuned CLIP ViT-L/14}
};
\node[modelbox] (prog) at (5.6,  0) {%
  \textbf{Progressive unfreezing}\\
  {\scriptsize fine-tuned CLIP ViT-L/14}
};
\node[modelbox] (zs)   at (5.6, -1.5) {%
  \textbf{Zero-shot}\\
  {\scriptsize base CLIP ViT-L/14}
};

\node[statsbox] (stats) at (0, -5.2) {%
  \centerline{\textbf{Statistical tests}}%
  \vspace{1pt}
  \centerline{\rule{42mm}{0.2pt}}%
  \vspace{3pt}
  \textbf{Zero-shot performance}\\
  ANOVA(facet)\\
  Spearman(depth)\\[3pt]
  \rule{42mm}{0.1pt}\\[2pt]
  \textbf{Fine-tuning gains}\\
  ANOVA(facet)\\
  Regression Analysis\\
  {\scriptsize $\Delta$NDCG@10 $\sim$ facet + depth}
};

\node[metricsbox] (metrics) at (5.6, -5.2) {%
  \textbf{Metrics}\\[1pt]
  \rule{34mm}{0.2pt}\\[4pt]
  $d_{\mathrm{mean}}$\\[3pt]
  \texttt{sim\_label}\\[3pt]
  Recall@$K$\\[3pt]
  NDCG@$K$
};


\draw[arr] (collections) -- (aat);

\coordinate (T1) at (2.9, 0);
\draw[edge] (collections.east) -- (T1 |- collections);
\draw[edge] (T1 |- lora) -- (T1 |- zs);                
\draw[arr]  (T1 |- lora) -- (lora.west);
\draw[arr]  (T1 |- prog) -- (prog.west);
\draw[arr]  (T1 |- zs)   -- (zs.west);

\begin{scope}[on background layer]
  \coordinate (T2) at ([yshift=-4mm]zs.south);
  \draw[edge] (zs.south)   -- (T2);
  \draw[arr]  (T2) -- (metrics.north);
\end{scope}

\draw[arr] (metrics.west) -- (stats.east);

\draw[arr] (aat) -- (stats);

\begin{scope}[on background layer]
  \node[groupbox, fit=(collections) (aat)] (g1) {};
  \node[groupbox, fit=(lora) (zs)]         (g2) {};
  \node[groupbox, fit=(stats) (metrics)]   (g3) {};
\end{scope}
\node[grouplbl] at (g1.north west) {Ground-truth data};
\node[grouplbl] at (g2.north west) {Models};
\node[grouplbl] at (g3.north west) {Analysis};

\end{tikzpicture}}%
\caption{Methodology overview. Ground-truth data (collections + AAT properties), three model conditions (zero-shot, LoRA, progressive unfreezing), and the analysis pipeline (metrics + statistical tests). Details in Section~\ref{sec:method}.}
\label{fig:methodology}
\end{figure}
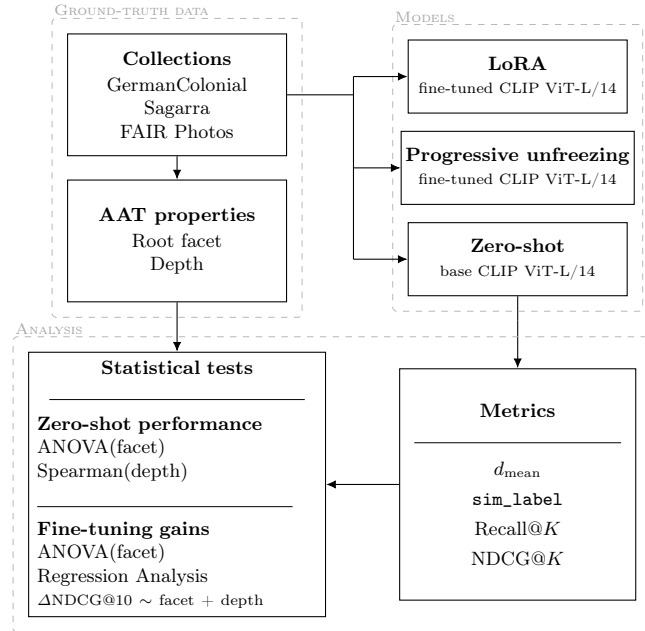

This section explains the extraction of AAT structural properties, CLIP fine-tuning and evaluation metrics (Figure~\ref{fig:methodology}).

\subsection{AAT structural properties}

Each photograph is associated with a set of AAT terms. 
For Sagarra and FAIR Photos these come from existing metadata. 
For GermanColonial labels in the metadata are connected to the AAT as described in Section~\ref{sec:GermanColonial}.
For each term, we extract two structural properties: the \textbf{root facet} (one of: \emph{Objects}, \emph{Activities}, \emph{Agents}, \emph{Materials}, \emph{Associated Concepts}, \emph{Styles and Periods}, \emph{Physical Attributes}, \emph{Brand Names}) and the \textbf{depth} (number of levels below the root facet).
\textbf{Root facet} aligns with the ontological type used in image classification systems~\cite{jorgensen2001}, and \textbf{depth} aligns with term abstraction, a known factor in CLIP performance~\cite{rodriguez2024,verma2023}.
Depth ranges from 1 to 13 across the three collections, with the majority of terms at depths 5--9 and the distribution of facets is shown in Table~\ref{tab:datasets}.
A term enters an analysis only if at least 20 images contribute to it: the zero-shot diagnostic analyses use terms with at least 20 images in the analysed collection, while the retrieval and fine-tuning-gain analyses use the terms covered by the held-out train/test split, which applies the same threshold to each term's images pooled across collections.

\subsection{Fine-tuning CLIP}

We evaluate two fine-tuning strategies applied to CLIP ViT-L/14 following previous work on fine-tuning CLIP for historical photographic collections~\cite{alijani2026}.
\textbf{Low-Rank Adaptation (LoRA)} ($r=16$, $\alpha=32$) applies low-rank adapters to all attention and feed-forward projection layers (\texttt{q\_proj}, \texttt{v\_proj}, \texttt{out\_proj}, \texttt{fc1}, \texttt{fc2}) and the visual and text projection heads, training for five epochs with a symmetric contrastive objective.
\textbf{Progressive unfreezing} trains in five phases, sequentially unfreezing transformer blocks from the top layer downward, with phase transitions driven by loss plateau detection, for approximately 45{,}000 steps total.
Meta-parameters for both strategies (LoRA rank, learning rate, epoch count, plateau thresholds) were set via a small number of informal preliminary runs rather than a systematic search on a separate validation split, and chosen to match the configuration reported by prior work on fine-tuning CLIP for historical photographic collections~\cite{alijani2026}.
During training, each photograph is paired with a text consisting of the term's preferred English label concatenated with a paraphrase of the term's AAT scope note.
Since AAT scope notes often exceed CLIP's 77-token context limit, they were paraphrased into single short descriptive sentences using \textit{Qwen3-8B}.
The paraphrase is used only as training-time augmentation. At query time we always use the AAT's preferred English label matching the task setting in prior work~\cite{alijani2026}.
Improvement due to fine-tuning ($\Delta$\,NDCG@10) is computed as the difference between the fine-tuned model and the zero-shot baseline on the held out 20\% test split.

\subsection{Metrics}

We compute four metrics per term.
Two that diagnose the source of retrieval difficulty and two that measure end-to-end retrieval performance.

\textbf{Visual coherence} ($d_{\text{mean}}$) is the mean pairwise cosine similarity of image embeddings within a term~\cite{leemann2022}.
Let $C$ denote the set of $n$ images assigned to an AAT term and $e_1,\dots,e_n$ their CLIP image embeddings:
\[
d_{\text{mean}}(C) = \frac{1}{\binom{n}{2}} \sum_{i < j} \cos(e_i, e_j)
\]
For AAT terms with more than 300 photographs we use a random sample of 300 for computational efficiency.

\textbf{Text-image centroid similarity} (\texttt{sim\_label}) is the cosine similarity between the CLIP text embedding of the term's preferred English label and the centroid of its image embeddings.
The zero-shot diagnostic analyses compute both metrics over all of a term's images in the analysed collection; comparisons between fine-tuned models and the zero-shot baseline compute all metrics on the held-out test split.

\textbf{Recall@$K$} and \textbf{NDCG@$K$} follow~\cite{alijani2026}, with the preferred English label as the query ranked against the shared pool of all test photographs.
Both retrieval metrics are computed at $K \in \{1, 5, 10\}$, and we report NDCG@10 in the paper body, since it correlates with NDCG@5 ($\rho=0.89$--$0.93$ across the zero-shot, LoRA, and progressive-unfreezing conditions) and NDCG@1 ($\rho=0.61$--$0.70$) as well as Recall ($\rho=0.78$--$0.80$).
Full results are given in the code repository accompanying this paper.\footnote{\url{https://github.com/rxvl-d/clip-aat-historical-photos-tpdl26}}

Together, these four measures capture different aspects of retrieval difficulty.
$d_{\text{mean}}$ reflects whether photographs cluster visually, independent of the text encoder and of category size.
\texttt{sim\_label} reflects whether the category label reaches that cluster, isolating the text-side failure mode.
Recall@$K$ and NDCG@$K$ measure end-to-end retrieval performance as a user would experience it, but conflate both failure modes and are sensitive to category size.
The two diagnostic measures allow us to distinguish between a category that retrieves poorly because its photographs have no visual structure and one where the structure exists but the label does not reach it.
Recall and NDCG also allow us to compare our results to previous work on CLIP retrieval in historical photographic collections~\cite{alijani2026} which only reports those metrics.

\subsection{Statistical analysis}

\paragraph{Facet and depth correlations.}
For zero-shot performance, one-way ANOVA tests whether root facet explains variance in $d_{\text{mean}}$ and \texttt{sim\_label}.
Effect sizes are reported as $\eta^2$.
Spearman rank correlation tests the relationship between depth and each metric.
Since deeper terms tend to have fewer images, we additionally compute partial Spearman correlations controlling for log image count.
All tests are run per collection and are two-tailed.
To check the monotonicity assumption behind Spearman $\rho$, we fit depth + depth$^2$ OLS models per collection per metric (6 tests), giving little evidence of curvature (all $p \geq 0.24$ after correction).
All reported $p$-values are unadjusted for multiple comparisons. Given the number of tests in Tables~\ref{tab:facet_anova} and~\ref{tab:depth_corr}, we anchor conclusions on the pooled, archive-controlled models and on patterns consistent in direction across collections, and treat isolated per-collection significances as descriptive.

\paragraph{Fine-tuning gains.}
For each fine-tuned model we compute the per-term gain $\Delta m = m_{\text{fine-tuned}} - m_{\text{zero-shot}}$, where $m$ is a metric value (NDCG@10, \texttt{sim\_label}, or $d_{\text{mean}}$) for a given AAT term.
Concept frequency $n_f$ is the number of train-split photographs for a given term.
We use $\log n_f$ throughout because concept frequency spans several orders of magnitude across AAT terms.
Three analyses are run on $\Delta$\,NDCG@10.

First, Spearman($\Delta$\,NDCG@10, $\log n_f$) tests whether concept frequency alone explains gains, following~\cite{alijani2026} who find that categories with more training photographs improve more after fine-tuning.

Second, one-way ANOVA($\Delta$\,NDCG@10 $\sim$ facet) with $\eta^2$ is run separately per collection to check whether the facet effect holds in each collection individually (an effect found only in one collection could reflect that collection's subject matter rather than a structural property of the AAT).

Third, ordinary least squares (OLS) regression $\Delta$\,NDCG@10 $\sim \log n_f + \text{depth} + \text{facet}$ tests whether facet and depth explain variation in gains beyond what concept frequency alone accounts for.
This is necessary because depth, facet, and frequency are correlated, so a bivariate correlation with gains could reflect the frequency effect rather than a structural one.

\section{Experimental Results}

We present our results in three stages, following our two research questions.
We first test whether root facet type and hierarchy depth correlate with visual coherence, text-image alignment, and end-to-end retrieval in the zero-shot setting (RQ1).
We then decompose retrieval failure into the three modes this joint distribution defines.
Finally, we test whether the same two properties predict fine-tuning improvement (RQ2).

\subsection{Zero-shot visual coherence and text-image alignment}
\label{sec:zeroshot}

RQ1 asks whether AAT root facet type and hierarchy depth correlate with visual coherence, text-image alignment, and end-to-end retrieval.
Across all three collections, in the zero-shot setting, the two structural properties predict visual coherence most strongly and text-image alignment and end-to-end retrieval less.

\paragraph{Visual coherence.}
Table~\ref{tab:facet_anova} shows that root facet type correlates with $d_{\text{mean}}$ in GermanColonial and Sagarra, but not in FAIR Photos, where the effect size is essentially zero ($\eta^2=0.008$).
In GermanColonial, the facet effect is attributable entirely to \emph{Activities}, a small facet ($n=9$), and in Sagarra entirely to \emph{Agents} ($n=18$). Excluding either facet renders the remaining facets statistically indistinguishable ($p=0.22$ and $p=0.36$, respectively), matching the pattern observed across all facets in FAIR Photos, where no facet functions as a comparable outlier.

Table~\ref{tab:depth_corr} shows that depth correlates positively with $d_{\text{mean}}$ in GermanColonial ($\rho=+0.26$), FAIR Photos ($\rho=+0.27$), and the pooled analysis, but is absent in Sagarra ($\rho=-0.01$).
Closer examination reveals that shallow terms in Sagarra are already clustering nearly as tightly as the deep ones (mean $d_{\text{mean}}$ differs by only 0.005 between terms with the top and bottom three depths in the distribution).
For GermanColonial and FAIR Photos the differences are 0.032 and 0.025 respectively.
This could be a collection-specific effect since Sagarra's photographs are all press photos from a single photographer, while the other two collections are aggregated from multiple sources.
There could also be a qualitative difference.
The shallow terms in the Sagarra collections are mostly \emph{Agents} (\emph{workers}, \emph{students}, \emph{scholars}, \emph{societies}) and a small cluster of terms relating to religion such as \emph{religious holidays}, \emph{religious art}, \emph{religions}.
Whereas in GermanColonial the shallow terms are mostly \emph{Objects} (\emph{portraits}, \emph{sculpture}, \emph{furniture}, \emph{tools}), and in FAIR Photos mostly \emph{Activities} (\emph{advertising}, \emph{exhibitions}, \emph{elections}, \emph{parades}).

We also examine these correlations for category-size effects but found that partial correlations controlling for log image count are essentially unchanged (pooled $\rho=+0.23$, GermanColonial $+0.24$, FAIR Photos $+0.21$).
Since GermanColonial contributes the most terms and is the most \emph{Objects}-heavy collection (81\% of terms against 70\% and 38\%), we additionally checked whether the pooled depth correlation is carried by \emph{Objects} terms.
It is not: pooled across collections, the correlation is stronger among non-\emph{Objects} terms ($\rho=+0.33$) than among \emph{Objects} terms ($\rho=+0.15$).

\paragraph{Text-image alignment.}
Both properties predict \texttt{sim\_label} far more weakly than they predict $d_{\text{mean}}$.
Facet type does not explain variance in \texttt{sim\_label} in any collection (all $p \geq 0.21$).
Depth correlates positively with \texttt{sim\_label} in all three collections, but more weakly than for $d_{\text{mean}}$ ($\rho \approx 0.07$--$0.22$).
Interestingly, the pattern is the reverse of the coherence result. 
Sagarra is the only collection where depth reliably correlated with \texttt{sim\_label}, rising from 0.206 (shallow) to 0.219 (mid) to 0.223 (deep), while GermanColonial (0.196 to 0.201) and FAIR Photos (0.216 to 0.214) are relatively flat.
In contrast the coherence result above, Sagarra's photographs are non-uniformly aligned, while GermanColonial's and FAIR Photos' are non-uniformly coherent but uniformly aligned.
No collection shows a significant depth correlation for both metrics at once.
Visual coherence and text-image alignment thus need to be examined together, which we do in Section~\ref{sec:failure}.

\paragraph{Retrieval metrics.}
Neither structural property carries over to end-to-end zero-shot retrieval. 
The facet effect on zero-shot NDCG@10 is not significant in any collection (Tables~\ref{tab:facet_anova}).
Depth is uncorrelated in the pooled analysis and significant only in Sagarra (Table~\ref{tab:depth_corr}). 
The Sagarra exception is essentially the same effect as from \texttt{sim\_label}. 
Controlling for \texttt{sim\_label}, the depth--NDCG@10 correlation becomes insignificant ($r=0.14$, $p=0.19$), confirming it is inherited from the alignment effect already reported above rather than a separate structural relationship.

\begin{table}[t]
\caption{One-way ANOVA with root facet as predictor ($F$, $p$, $\eta^2$), for zero-shot metrics and fine-tuning gains ($\Delta$). 
\textbf{Bold} $\Rightarrow p<0.05$.
\textbf{Pooled} $\Rightarrow$ All collections together.
}
\label{tab:facet_anova}
\centering\scriptsize
\setlength{\tabcolsep}{2.5pt}
\resizebox{\textwidth}{!}{%
\begin{tabular}{l|l|rrr|rrr|rrr|rrr}
\toprule
& & \multicolumn{3}{c|}{Pooled} & \multicolumn{3}{c|}{GermanColonial} & \multicolumn{3}{c|}{Sagarra} & \multicolumn{3}{c}{FAIR Photos} \\
\cmidrule(lr){3-5}\cmidrule(lr){6-8}\cmidrule(lr){9-11}\cmidrule(lr){12-14}
Variable & Setting & $F$ & $p$ & $\eta^2$ & $F$ & $p$ & $\eta^2$ & $F$ & $p$ & $\eta^2$ & $F$ & $p$ & $\eta^2$ \\
\midrule
\multirow{3}{*}{$d_{\text{mean}}$} & zero-shot & \textbf{2.76} & \textbf{0.012} & \textbf{0.029} & \textbf{4.44} & \textbf{$<$0.001} & \textbf{0.056} & \textbf{3.54} & \textbf{0.017} & \textbf{0.090} & 0.21 & 0.890 & 0.008 \\
 & $\Delta$ LoRA & 0.52 & 0.796 & 0.006 & 0.93 & 0.464 & 0.011 & 0.19 & 0.904 & 0.007 & 0.60 & 0.618 & 0.027 \\
 & $\Delta$ progressive & 1.75 & 0.107 & 0.019 & 1.65 & 0.145 & 0.020 & 1.97 & 0.125 & 0.069 & 0.41 & 0.743 & 0.019 \\
\multirow{3}{*}{\texttt{sim\_label}} & zero-shot & 0.74 & 0.620 & 0.008 & 0.78 & 0.561 & 0.010 & 1.51 & 0.216 & 0.041 & 1.27 & 0.289 & 0.046 \\
 & $\Delta$ LoRA & 1.33 & 0.242 & 0.014 & 0.77 & 0.569 & 0.010 & \textbf{4.94} & \textbf{0.003} & \textbf{0.156} & 0.60 & 0.615 & 0.027 \\
 & $\Delta$ progressive & \textbf{2.72} & \textbf{0.013} & \textbf{0.029} & 1.40 & 0.222 & 0.017 & \textbf{7.74} & \textbf{$<$0.001} & \textbf{0.225} & 0.73 & 0.539 & 0.032 \\
\multirow{3}{*}{NDCG@10} & zero-shot & 1.94 & 0.073 & 0.021 & 1.07 & 0.374 & 0.013 & 1.93 & 0.131 & 0.068 & 1.43 & 0.241 & 0.062 \\
 & $\Delta$ LoRA & 1.78 & 0.101 & 0.019 & \textbf{2.86} & \textbf{0.015} & \textbf{0.034} & 0.30 & 0.828 & 0.011 & 0.08 & 0.970 & 0.004 \\
 & $\Delta$ progressive & 1.89 & 0.080 & 0.020 & 2.08 & 0.067 & 0.025 & 2.30 & 0.084 & 0.079 & 1.62 & 0.194 & 0.070 \\
\bottomrule
\end{tabular}}
\end{table}

\begin{table}[t]
\caption{Spearman $\rho$ between AAT hierarchy depth and each correlated variable. In the case of the fine-tuning strategies, the variable is the $\Delta$ over the zero-shot metrics. \textbf{Bold} $\Rightarrow p<0.05$.}
\label{tab:depth_corr}
\centering\scriptsize
\setlength{\tabcolsep}{3pt}
\begin{tabular}{l|l|rr|rr|rr|rr}
\toprule
& & \multicolumn{2}{c|}{Pooled} & \multicolumn{2}{c|}{GermanColonial} & \multicolumn{2}{c|}{Sagarra} & \multicolumn{2}{c}{FAIR Photos} \\
\cmidrule(lr){3-4}\cmidrule(lr){5-6}\cmidrule(lr){7-8}\cmidrule(lr){9-10}
Variable & Setting & $\rho$ & $p$ & $\rho$ & $p$ & $\rho$ & $p$ & $\rho$ & $p$ \\
\midrule
\multirow{3}{*}{$d_{\text{mean}}$} & zero-shot & \textbf{+0.249} & \textbf{$<$0.001} & \textbf{+0.261} & \textbf{$<$0.001} & -0.007 & 0.938 & \textbf{+0.269} & \textbf{0.013} \\
 & $\Delta$ LoRA & +0.082 & 0.050 & +0.024 & 0.634 & +0.065 & 0.558 & -0.116 & 0.336 \\
 & $\Delta$ progressive & \textbf{+0.150} & \textbf{$<$0.001} & +0.025 & 0.616 & +0.205 & 0.061 & +0.010 & 0.935 \\
\multirow{3}{*}{\texttt{sim\_label}} & zero-shot & +0.037 & 0.390 & +0.069 & 0.181 & \textbf{+0.218} & \textbf{0.021} & +0.190 & 0.082 \\
 & $\Delta$ LoRA & \textbf{-0.164} & \textbf{$<$0.001} & \textbf{-0.117} & \textbf{0.018} & \textbf{-0.221} & \textbf{0.043} & -0.168 & 0.161 \\
 & $\Delta$ progressive & \textbf{-0.166} & \textbf{$<$0.001} & \textbf{-0.098} & \textbf{0.048} & \textbf{-0.239} & \textbf{0.028} & -0.224 & 0.060 \\
\multirow{3}{*}{NDCG@10} & zero-shot & +0.003 & 0.936 & +0.016 & 0.747 & \textbf{+0.244} & \textbf{0.025} & +0.154 & 0.200 \\
 & $\Delta$ LoRA & \textbf{-0.100} & \textbf{0.018} & -0.067 & 0.176 & +0.110 & 0.321 & -0.079 & 0.513 \\
 & $\Delta$ progressive & \textbf{-0.134} & \textbf{0.001} & -0.004 & 0.931 & -0.086 & 0.434 & \textbf{-0.252} & \textbf{0.034} \\
\bottomrule
\end{tabular}
\end{table}

\subsection{Analysis of failure modes}
\label{sec:failure}

\begin{figure}[H]
\centering
\includegraphics[width=0.9\textwidth]{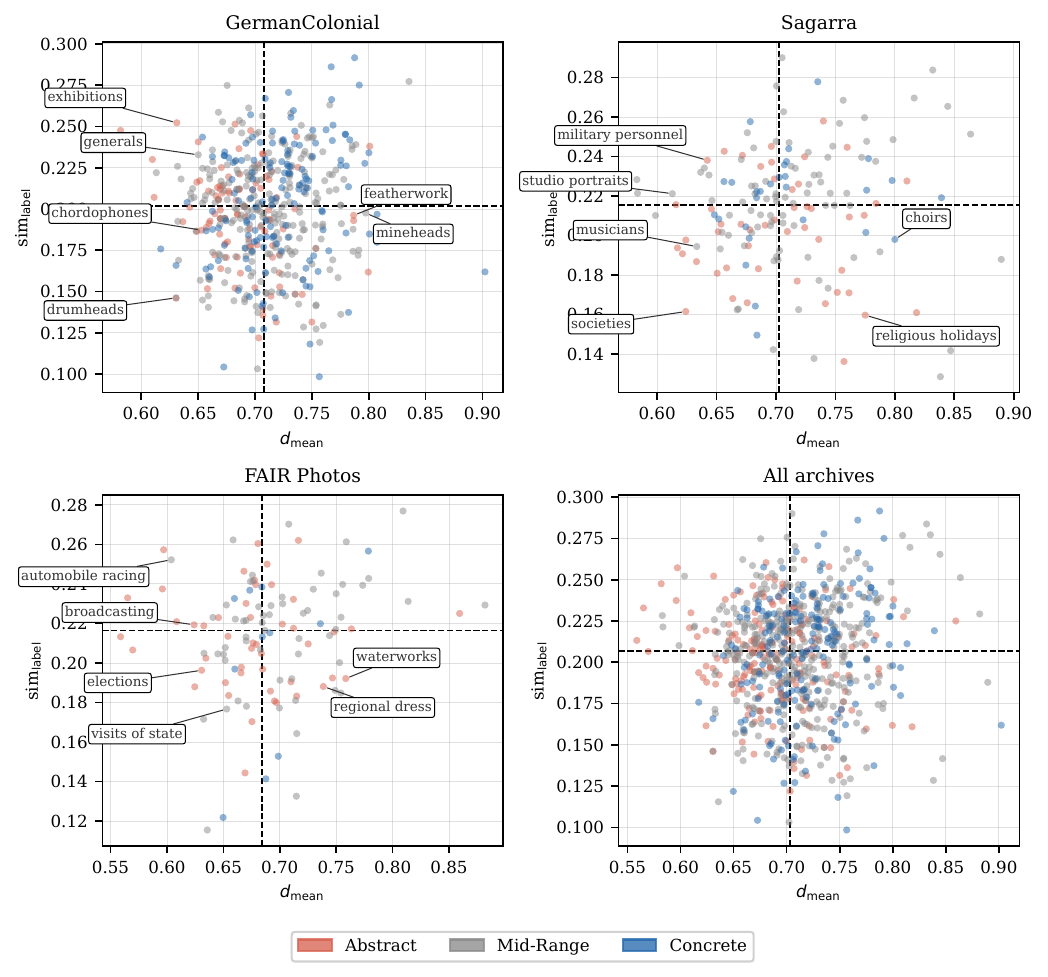}
\caption{Joint distribution of $d_{\text{mean}}$ and \texttt{sim\_label}.
\textit{Abstract} (depth $\leq 5$), \textit{Mid-Range} (depth $6$--$8$), \textit{Concrete} (depth $\geq 9$).
Dashed lines mark the per-collection medians. Annotated examples illustrate the three failure modes described in Section~\ref{sec:failure}.
}
\label{fig:quadrant}
\end{figure}

The joint distribution of $d_{\text{mean}}$ and \texttt{sim\_label} shows the three failure modes (Fig.~\ref{fig:quadrant}).
Each term falls into one of four quadrants, split at the per-collection medians of the two metrics, abbreviated by whether $d_{\text{mean}}$ and \texttt{sim\_label} are high or low (HH, HL, LH, LL).
We would expect retrieval success to occur when a term's images form a tight cluster and the label lands near it (i.e. the HH quadrant).
The other three are the three failure modes:

\textbf{High $d_{\text{mean}}$, low \texttt{sim\_label} (HL).} Images cluster visually but the label lands far from the cluster.
This quadrant contains both domain-specific object vocabulary (\textit{featherwork}, \textit{mineheads}, \textit{fishing tackle}) and activity terms (\textit{religious holidays}).
The \emph{Materials} facet illustrates this pattern at the facet level. 
Pooled across collections, its zero-shot $d_{\text{mean}}$ (0.716) is the highest of any facet, while its \texttt{sim\_label} (0.192) is the lowest.

\textbf{Low $d_{\text{mean}}$, high \texttt{sim\_label} (LH).} The label lands near the image centroid but the photographs scatter.
This quadrant contains broad agent categories (\textit{generals}, \textit{military personnel}, \textit{studio portraits}) and activity terms (\textit{exhibitions}, \textit{automobile racing}).
The problem here is the visual heterogeneity of the category rather than the text encoder.

\textbf{Low $d_{\text{mean}}$, low \texttt{sim\_label} (LL).} Neither photographs cluster nor does the label reach them.
The LL quadrant captures AAT terms ranging from highly domain-specific object vocabulary (\textit{drumheads}) to broad activity categories (\textit{elections}, \textit{societies}, \textit{broadcasting}).

The two metrics are nearly uncorrelated pooled across collections (Spearman $\rho=-0.03$, $p=0.53$) and at most weakly correlated within them ($\rho \leq +0.25$).
Zero-shot NDCG@10 differs by quadrant.
In GermanColonial and Sagarra, HL retrieves worst of all four quadrants (0.027 and 0.034), below LL (0.051 and 0.044).
In FAIR Photos, HL (0.162) is slightly better than LL (0.114).
LH is worse than HH in Sagarra (0.071 vs.\ 0.233) and FAIR Photos (0.315 vs.\ 0.444), but not in GermanColonial (0.126 vs.\ 0.144)
(Full per-quadrant counts and NDCG@10 are in the previously linked repository).

That a category can retrieve worse with one intact metric can be explained by the retrieval mechanism. 
A tight cluster that sits far from its label misses the top ranks uniformly, while the scattered photographs of an LL term still place a few images near the label by chance.
A visually coherent but unaligned cluster thus manifests as a wrong ranking as opposed to an accidentally right one.

\subsection{Fine-tuning gains}

Tables~\ref{tab:facet_anova} and~\ref{tab:depth_corr} report the facet and depth effects on NDCG@10 gains under fine-tuning.
Zero-shot NDCG@10 averages 0.114 across 563 AAT terms.
LoRA fine-tuning raises this to 0.252 ($\Delta=+0.138$) and progressive unfreezing to 0.259 ($\Delta=+0.145$).

\paragraph{Alignment gains vs.\ coherence gains.}
Depth predicts gains in \texttt{sim\_label} but not in $d_{\text{mean}}$. 
Shallower AAT terms improve alignment more under both LoRA ($\rho=-0.164$, $p<0.001$) and progressive unfreezing ($\rho=-0.166$, $p<0.001$).
Facet explains no coherence-gain variance in any collection ($p \geq 0.107$ throughout). Depth's coherence-gain correlation is significant only across collections and only for progressive unfreezing ($\rho=+0.150$).
Fine-tuning thus mostly improves text-image alignment where it was weakest (i.e. in shallower terms), with no comparable predictable effect on visual coherence.

\paragraph{Frequency vs.\ facet as predictors.}
Concept frequency ($\log n_f$) correlates with NDCG@10 gains for LoRA ($\rho=+0.217$, $p<0.001$) but not for progressive unfreezing ($\rho=+0.027$, $p=0.529$), replicating~\cite{alijani2026} only partially.
Controlling for frequency in an OLS model ($\Delta\,\text{NDCG@10} \sim \log n_f + \text{depth} + \text{facet}$), the facet joint F-test is significant for both LoRA ($F=5.67$, $p<0.001$) and progressive unfreezing ($F=8.55$, $p<0.001$; $R^2=0.098$ and $0.104$ respectively). 
This means that facet explains gain variance that frequency alone does not.
This holds specifically for alignment. 
The facet effect on $\Delta$\,\texttt{sim\_label} is significant when pooled across collections for progressive unfreezing ($p=0.013$) and within Sagarra under both strategies (Table~\ref{tab:facet_anova}).
The size of the effect is collection-dependent, the facet $\times$ collection interaction is significant for progressive unfreezing ($F=2.25$, $p=0.029$).
Controlling for collection weakens the facet effect on $\Delta$\,NDCG@10 to marginal for both strategies ($F=1.78$, $p=0.10$; $F=1.89$, $p=0.08$).

\section{Discussion}

The central finding of this study is that visual coherence and text-image alignment are nearly uncorrelated across AAT terms (Section~\ref{sec:failure}), so a retrieval score such as NDCG@10, which depends on both, does not show which of the two failed.
In answer to RQ1, root facet and depth correlate with visual coherence, each significantly in two of three collections and in the pooled analysis.
Their correlation with text-image alignment is weak and inconsistent, and neither property correlates with retrieval performance at all.
This follows from the near-zero correlation between the two metrics. A structural property that predicts only one of them shows little relationship with a score that depends on both.

Low alignment is the more consistent failure factor across collections: HL terms rank worst or second-worst of the four quadrants everywhere, often below LL terms that fail on both metrics, since a mislabeled tight cluster ranks consistently low (Section~\ref{sec:failure}).
Earlier accounts that explain variance in Recall@K through abstraction level~\cite{rodriguez2024} or concept frequency~\cite{parashar2024,alijani2026} therefore mix failure modes with distinct causes and, as argued below, distinct remedies.

The facet effect anticipated by LIS frameworks that treat ontological type as a primary axis of image content~\cite{jorgensen2001,westman2009} is not strongly present here, and holds in each collection only through a single facet (\emph{Activities} in GermanColonial, \emph{Agents} in Sagarra; Section~\ref{sec:zeroshot}).
The depth effect is more robust, but its absence in Sagarra suggests visual coherence is determined jointly by vocabulary granularity and collection composition, not by the vocabulary alone.

Text-image alignment shows the opposite pattern. In the zero-shot setting, neither facet nor depth predicts it, yet alignment is what fine-tuning changes, improving most for shallow terms where it was weakest, without affecting coherence.
Fine-tuning with an image-text contrastive objective therefore addresses the coherent-but-misaligned (HL) failure mode but not visual scatter (LH).

At the observed effect sizes, neither facet nor depth reliably predicts retrieval behaviour for an individual AAT term.
They can instead point to parts of a vocabulary that are likely to contain retrieval problems (e.g.\ shallow, broad terms), whose terms are then evaluated with the two diagnostic metrics, and the resulting failure mode indicates a response.
For visually incoherent categories, retrieval-agnostic access (faceted browsing, seed-image similarity search) avoids relying on a label to retrieve a cluster that does not exist.
For coherent but misaligned categories, domain adaptation can move the label embedding toward the existing cluster, as the fine-tuning results indicate, and query expansion may identify phrasings that lie closer to the image cluster than the archival term.
For categories whose label is recognised but whose photographs scatter, the standard image-text contrastive loss is unlikely to help, since it pulls images toward a label embedding that is already near the centroid without reducing scatter.
A supervised contrastive objective over image embeddings~\cite{khosla2020}, treating photographs of the same AAT term as positives, would instead reduce intra-category scatter and increase $d_{\text{mean}}$.

\paragraph{Limitations and Future Work.}
As the effect sizes above show, the structural properties explain only modest variance, and the OLS models for gain prediction reach $R^2 \approx 0.10$, meaning most of the variation across AAT terms remains unaccounted for.
The GermanColonial links are automatically derived and cover 56\% of images and 17.6\% of the collection's keyword vocabulary.
The mapped subset is skewed toward concepts with clean German-to-AAT string matches, which plausibly also over-represents concepts in CLIP's training distribution.
We plan to investigate the suggested practical responses to each failure mode, and whether they are effective in improving retrieval.

\section{Conclusion}

We investigated whether AAT root facet type and hierarchy depth explain where CLIP retrieval fails and where fine-tuning helps, across three historical photographic collections.
Visual coherence and text-image alignment are nearly uncorrelated across terms, so a single retrieval score conflates failure modes with distinct causes, and the worst-performing terms are often those that are coherent but misaligned rather than those failing on both metrics.
Vocabulary structure predicts coherence but not alignment or zero-shot retrieval, while fine-tuning improves alignment for shallow, broad terms and leaves coherence largely unchanged.
These results point toward failure-mode-specific remedies: alignment fixes such as query expansion for coherent-but-misaligned terms and contrastive training for visually scattered ones, as the next step toward improving retrieval.

\begin{credits}
\subsubsection{\ackname}
This work was supported by the Deutsche Forschungsgemeinschaft (DFG, German Research Foundation), project number: 531013489 (\href{https://vabiko.eu/}{VABiKo}).

\subsubsection{\discintname}
The authors have no competing interests to declare.
\end{credits}

\bibliographystyle{splncs04}
\bibliography{references}

\end{document}